# *Sarong* rolled around a body demonstrates the force for separating two sheets joined by folding and rolling is very large

Fathan Akbar and Mikrajuddin Abdullah[a]

Department of Physics, Bandung Institute of Technology

Jl. Ganesa 10 Bandung 40132, Indonesia

[a]Email: mikrajuddin@gmail.com

Abstract

There are a lot of new sciences that have been inspired by common phenomena and even by old traditions practiced in our daily lives. Eventually, it may induce unexpected new technologies after deeply explored. In this paper, inspired by the wearing of traditional cloth, named *sarong*, by the community in the South East Asian countries and others, we investigated the behavior of sheets folded like rolling of *sarong* around the stomach. Simple equipment was designed to qualitatively collect the data and combined with simple modeling. Rolling of the *sarong* around the stomach generates the joining force of two sheets, increasing proportionally to the square of the number of rolling. This finding is potentially applied for developing a method for strongly joining sheets by simply rolling and releasing the join by simply unrolling. This work can also be simply duplicated elsewhere so that it is worthy of teaching materials at undergraduate levels. Both scientific and teaching contents can be extracted simultaneously.

# 1. Introduction

Origami science opens opportunities for the development of technologies in the future by employing very efficient materials. The heart of origami science is the folding of sheets. Today we can see many buildings, the architecture of which has been inspired by origami structure, such as Festival Hall of the Tiroler Festspiele Erl, Austria, Klein Bottle House Rye, Victoria, Australia, Park Pavilion Cuenca, Spain, and Tel Aviv Museum of Art, Israel [1].

Currently, scientists are considering a new technology for exploring exoplanets by designing a very large starshade having the origami structure. This umbrella-like module will be placed in the space at a distance of about 77,000 km in front of the optical telescope. This module will block the bright light from star to enter the telescope and only allows dim light from exoplanets revolving the center star. This module must be transported into space in a small size and opened when it reaches the desired position. Using the origami structure, it can be transported in a folded structure and then unfolded after reaching the right position in the space [2-5].

Although there have been a lot of research in folded structures, however, new topics will always appear. It sometimes happens these new topics are inspired by common phenomena in our daily life. At present, we will investigate the behavior of folded sheets, the folding geometry of which was rarely been reported. This idea was inspired by the traditional cloth in South East Asian countries or others, where many people still use *sarong* in their daily lives.

*Sarong* is made of one sheet of fabrics and sewn like an easily-folded cylindrical wall. When worn, the *sarong* is simply wrapped around the stomach without using any belts or other objects as hooks. It is firmly attached to the stomach after rolling the front part several times. Having rolling for about five times, the sarong is hardly released by pulling forces. We are going to explore this behavior and expect to identify new physics ingredients and suggesting a potential new mechanism for tightly joining and separating sheets just by rolling and unrolling.

At first glance, this topic seems very "traditional". But, indeed, there are many "traditional" topics have been explored by authors and they might initiate new technologies.

For example, Reis's group from MIT has investigated many common phenomena in our daily life, and they have demonstrated many exotic findings [6-11]. Deegan et al have investigated the formation of coffee stain ring when the coffee drop is dried [12]. Goldberg and O'Reilly described the geometry evolution of cooking spaghetti [13]. The present author has also investigated simple common phenomena such as bending of fireworks [14], wringing of wet cloth [15], rolling of a cylinder containing granules [16,17], sand tunnel collapse [18], rice winnowing [19], measuring the atmospheric temperature in an aircraft cabin [20], and bending of vertical sheets that lead to a phase transition [21].

In this work, we designed a simple experiment method so that it can be duplicated elsewhere by teachers of undergraduate students. The experiment tools were made of simple utensils. We believe that this method is worthy of teaching materials in undergraduate courses, especially for schools or universities those are not facilitated with standard research equipment.

## 2. Experiment

First, we will estimate the pressure generated in the location between the stomach and the rolled part of the *sarong* as the rolling number of the *sarong* is increased. We designed a simple tool as shown in Fig. 1 by employing the principle of hydrostatic pressure. We used a flexible transparent hose, bent like a U-letter, and filled with colored water. One end of the hose was open and the other end was connected to a toy balloon containing air at 1 atm pressure. Figure 1(b) shows the complete measuring system.

The pressure was measured by firstly rolling the *sarong* around the stomach (see Appendix). After making three rollings, the condition where the *sarong* has been wrapped tightly around the stomach, the balloon was then inserted in the position between the *sarong* and the stomach, causing the water level inside the hose to change. Due to breathing, the water level changed constantly. We decided to record the highest difference in the water level. In addition, we assume the balloon membrane did not differentiate the pressure of the air inside the balloon and the pressure between the stomach and the *sarong* (assuming the balloon membrane is very flexible).

We also measured the friction forces of various sheets for comparison. We measured the force required to move horizontally a stack of sheets that were arranged alternately. The movable sheets were clamped together to a movable clamp. This clamp was attached to a flexible rope passing through a pulley to a load. The fixed sheets were clamped together to a fixed clamp. The sheet stack was placed between two glass plates, each of which has been fixed to a wooden block. A pressure load made of a plastic container, inside of which water was filled in, was placed above the upper wooden block. This permits the change of mass easily just by changing the water volume. Figure 2 is an illustration of the measurement tool.

The rope supports a forcing (hanging) load made of a plastic container, inside of which water was filled in too. At each pressure load mass, the water is added gently into the forcing load container. We record the mass of the forcing load when the movable clamp starts to move. We used photocopy paper sheets, envelope paper sheets, and a wool fabric. For these two paper sheets, we measured up to five movable sheets (until eleven sheets were arranged in the stack), while for the wool fabric we only measured up to four movable sheets (since it is thicker).

In the stack, the bottom and the top sheets were the fixed sheets so that the movable sheet only contact with the fixed sheet, not with the glass. Inset in Fig. 2 is the position of the sheets. Yellow color represents the movable sheets and violet color represents the fixed sheets. We varied the mass of the pressure load (the total mass after adding the mass of upper grass plate and wooden block): 0.75 kg, 1.00 kg, 1.25 kg, and 1.50 kg.

To mimic the rolled part in a *sarong* that wrapped around the stomach, we measured the force required to release folded sheets. We used also the equipment in Fig. 2. They were different in the arrangement of the sheet between two glass plates. Instead of using a stack arrangement, we placed a folded sheet between two fixed sheets. One end of the folded sheet was attached to the movable clamp and the other end was attached to the fixed clamp. The bottom and the top layers are the fixed sheets so that the folded sheet contact only with the fixed sheets, not with the glass, as illustrated in Fig. 3(a).

The folding geometry is shown in Fig. 3(b). We generated two types of folding: cross folding and side folding. Figure 3(b) represents only the cross folding without side folding. The cross folding and one side folding are shown in Fig. 3(c). The sequence from (c1) to (c3) is the process of making one side folding. A sheet having double width is firstly cross-folded as in making a zero side-folding (the same as Fig 3(b) with double width). This sheet is then side-folded at the center and the final result is shown in Fid. 3(c). This geometry might be compared to rolling the *sarong* once. To make a twice side-folding, we used a sheet having triple width. We repeated the process as in (c1) to (c3) by making two times side-folding.

## 3. Results and Discussion

Figure 4 shows the effect of the number of rolling on the pressure between the rolled part of the *sarong* and the stomach. The pressure was calculated using the hydrostatic pressure equation $\Delta P = \rho g h$, with $\rho$ = 1,000 kg/m$^3$ is the water density (the colored substance did not change the water density since the amount was very small), $g$ is the acceleration of gravitation, and $h$ is the difference in the water level in the hose. We used $g$ = 9.77 m/s$^2$, the value for Bandung city, Indonesia, as reported by Khairurrijal et al [22]. We see that the pressure, $\Delta P$, increased linearly with the rolling number of the *sarong*, $n$, with the best fitting of

$$\Delta P = 0.7n \tag{1}$$

This behavior can be explained as following. We assume the compression of the stomach surface behaves elastically. By increasing the number of rolling the sarong from $n$ to $n+1$, the stomach surface will displace inward as $\Delta x$. Therefore, an additional elastic force is generated by the stomach to the *sarong* as $\Delta F = k \Delta x$, with $k$ is the spring constant. This relationship leads to the elastic force equation $F = kx \propto n$. Again, by assuming the contact area between the *sarong* and the stomach nearly unchanged, we obtain $P = F/A \propto n$, which is consistent with data in Fig. 4.

From this result, we found that increasing the number of rolling, implied the increase in pressure and the total contact area. Suppose the initial radius of the rolling is $r_0$ and the thickness of one rolling sheet is $\Delta x$ (generally thicker than the sheet thickness due slightly wrapping). The rate change in the radius becomes $\Delta r / \Delta \theta = \Delta x / 2\pi$ with $\theta$ is the rolling angle. Therefore, the radius after rolling by an angle $\theta$ is

$$r = r_0 + \frac{\Delta x}{2\pi} \theta \tag{2}$$

When the rolling angle changes by $d\theta$, the arc length changes by $ds = rd\theta = (r_0 + (\Delta x/2\pi)\theta)d\theta$. If the effective width of the rolled *sarong* is $w_{ef}$, the total contact area after $n$ complete rollings is

$$A = \int_0^{2n\pi} w_{ef} ds = 2n\pi r_0 w_{ef}\left(1 + \frac{1}{2}\frac{\Delta x}{r_0}n\right) \qquad (3)$$

The number of rolling is generally less than ten ($n \leq 10$). The *sarong* thickness is around one millimeter, while the initial radius might be several centimeters. Therefore, in general, $n\Delta x/2r_0 << 1$ and we obtained an approximated area $A = 2\pi r_0 w_{ef} n$. Rolling of the *sarong* increases the pressure and the area, each of which is proportional to the number of rolling. The force that compresses the *sarong* sheet is $f = \Delta PA \propto n^2$. This force plays a role as a normal force in case of friction between contacting surfaces so that the force required to release the rolled part of the *sarong* satisfies

$$F = \mu f \propto \mu n^2 \qquad (4)$$

with $\mu$ is the "friction coefficient" and might depend on the number of rolling, $\mu(n)$. For merely touching surfaces, such as a block placed on a surface, the friction coefficient can be considered to be constant. Although, there is also a proposal that even between merely touching surfaces, the friction coefficient might depend on the normal force. For example, Konecny reported the friction force satisfies $N^{0.91}$ to mean that $\mu \propto N^{-0.09}$ with $N$ is the normal force [23]. We will confirm Eq. (4) by simple experiments. For this demonstration, we used paper and fabric sheets to show that the behavior applies for general sheets, instead of fabrics only.

First, we determine the friction of sheets without folding. The sheets are arranged in the stack, as shown in the inset of Fig. 2. Figure 5 is the measurement data for (a) photocopy paper sheets, (b) envelope paper sheets, and (c) wool fabric sheets. We varied the pressure load placed above the upper glass: 0.75 kg, 1.00 kg, 1.25 kg, and 1.50 kg. The forcing load mass is increased gently until the movable clamp (sheets) starts to move. We obtained that all data show a linear increase in forcing load when increasing the number of sheets. We also observed the slope increased with increasing the pressing mass.

The contact area is proportional to the number of sheets, or $A \propto n$. From all data, we can deduce a relationship between the forcing load, $W$, and the number of sheets as

$$W = \Omega(M_p) n \tag{5}$$

with $\Omega(M_p)$ is a factor which depends on the pressing mass, $M_p$. Based on all data in Fig. 5, it is very clear that $\Omega(M_p)$ increases with $M_p$. Our next objective is to determine the dependence of $\Omega(M_p)$ on $M_p$.

From Eq. (5) that $\Omega(M_p)$ is the slope of the curve $W$ against $n$. To determine the dependence of $\Omega(M_p)$ on $M_p$ we plot $\Omega(M_p)$ against $M_p$ and then searching for the fitting curve. The $\Omega(M_p)$ was obtained from the fitting curves in Fig. 5. For the photocopy paper sheets and the envelope paper sheets, we identified $\Omega(M_p) = cM_p$ with $c$ is a constant. However, for the wool fabric sheets, we obtained $\Omega(M_p) = c_1 M_p + c_2$ with $c_1$ and $c_2$ are constants. This difference possibly caused by the fact that the wool fabric is much thicker than the paper sheets and the fabric sheet sometimes is not flat perfectly. In addition, the wool fabric surface is likely hairy. Based on this evidence, we may claim that for smooth sheets, the relationship $\Omega(M_p) = cM_p$ is satisfied. Combining with Eq. (5) we then obtain

$$W = cM_p n \tag{6}$$

Now let us investigate the effect of side-folding on the forcing load. This side-folding mimics the roll of the *sarong*. We measured for the photocopy paper sheets and the envelope paper sheets. Figure 6(a) is the data for photocopy paper sheets and 6(b) is the data for envelope paper sheets. We fixed the pressing load mass of 1.50 kg.

If we plot in a linear scale, we obtain nonlinear curves. Therefore, we tried a different scale to obtain a linear change. We identified that the linear change was obtained if we plot the load against the square of the number of side-folding. Surprisingly, the slope of the curve belongs to the photocopy paper sheets (1.092), and the envelope paper sheets (1.163) are nearly the same, and two curves nearly coincide. From this data, we then obtain the relationship between the load and the number of side-folding as the $W = a + bn^2$, with $a$ and $b$ are constants. From this equation we can write

$$W = b\left(1 + \frac{a}{bn^2}\right)n^2 \qquad (7)$$

By comparing Eq. (4) and (7) we obtain the friction coefficient changes according to $\mu \propto (1 + \gamma/n^2)$ with $\gamma$ is a constant. Until these results, we can claim that the force to release the sarong is identical to the force for releasing the folded sheets and it increases according to the square of the rolling number.

We observed the loads for releasing the sheet stack is different from the load for releasing the side folding sheets. The later force increased rapidly than the previous one. This is due to the generation of the fording edges. There are two folding edges that are created: the cross folding edge and the side-folding edge. For zero side-folding as shown in Fig. 3(b), only a cross folding edge exists. However, for geometry containing side-folding, the side-folding edge is also generated.

To create a folding edge, we need a certain force. This is similar to the force required for producing crease [24-26]. Moving the folding sheet is identical to generating a new crease so that need additional force compared to moving sheet without folding edges as in the stack structure is required. Edler et al described that the force required to fold a film scales as $F \propto bB/H^2$, with $b$ is the width of the film, $H$ is the size of the gap between two films after folding, and $B$ is the bending modulus [27].

Finally, we investigated the effect of the width of the folded paper sheet (photocopy paper) on the forcing load. We changed the width from 0.01 m to 0.07 m without side-folding (inset (a) in the figure). The results are shown in Fig. 7, square symbols. The load increased with the width of the sheet. Fitting the data with a linear curve, we obtain a function $W = 23.4t + 1.026$ (kg). This equation might be less accurate for very small width since it must be $W \to 0$ as $t \to 0$. The curve predicts the load becomes $W = 3.366$ kg when the sheet width becomes 0.1 m (point A in the figure).

We also investigated the effect of asymmetrical folding size on the forcing load. Paper sheets of 0.1 m width were folded at different width ratios: 2.5/7.5, 4/6, and 5/5. Inset b in Fig. 7 is the description of the folding geometry. Circle symbols in Fig. 7 are the measured results. Fitting the data with a linear function resulted in an approximated equation $W = 0.37(t_1/t_2) + 3.255$ (kg). It is clear that the change of the load with the ratio of ($t_1/t_2$) is so weak. One interesting result is when $t_1 = 0$ or $t_1/t_2 = 0$ we have the curve cross the load axis at 3.255 kg (see point B in Fig. 7).

Indeed, points A and B in Fig 7 represent the same condition. Point A is obtained when the size of the sheet (without side folding) is 0.1 m. Point B is the condition when $t_1 = 0$ and $t_2 = 0.1$ m (the condition of without folding). Therefore both points are identical and have been obtained from extrapolation of two different measurements. Their values are nearly identical (3.366 kg for point A and 3.255 kg for point B).

## 4. Conclusion

We have been able to design a simple tool for extracting the mechanical behavior of wearing the traditional cloth, named *sarong*, by rolling around the stomach. Rolling generates very high joining forces that increase with the square of rolling numbers. This type of joining is very interesting since we do not need a kind of belt or another hook object. Data collected from the experiment using paper sheets and wool fabric sheets were consistent with a simple model. This topic sounds very simple, but there are many proves, even common phenomena around us can generate new sciences and new technology.

## References

[1] https://architizer.com/, accessed on May 26, 2020.

[2] Carter J, A Giant Origami Space Telescope That Can Photograph Alien Worlds. Is This NASA's Next Flagship Mission?, 2019, December 18 Forbes, accessed on May 26, 2020.

[3] Amos J, James Webb: Telescope's giant origami shield takes shape, 2017 August 9, BBC.com, accessed on May 26, 2020.

[4] Wilson L, Pellegrino S, Danner R, Origami Sunshield Concepts for Space Telescopes, http://www.its.caltech.edu//~sslab/PUBLICATIONS, accessed on May 26, 2020.

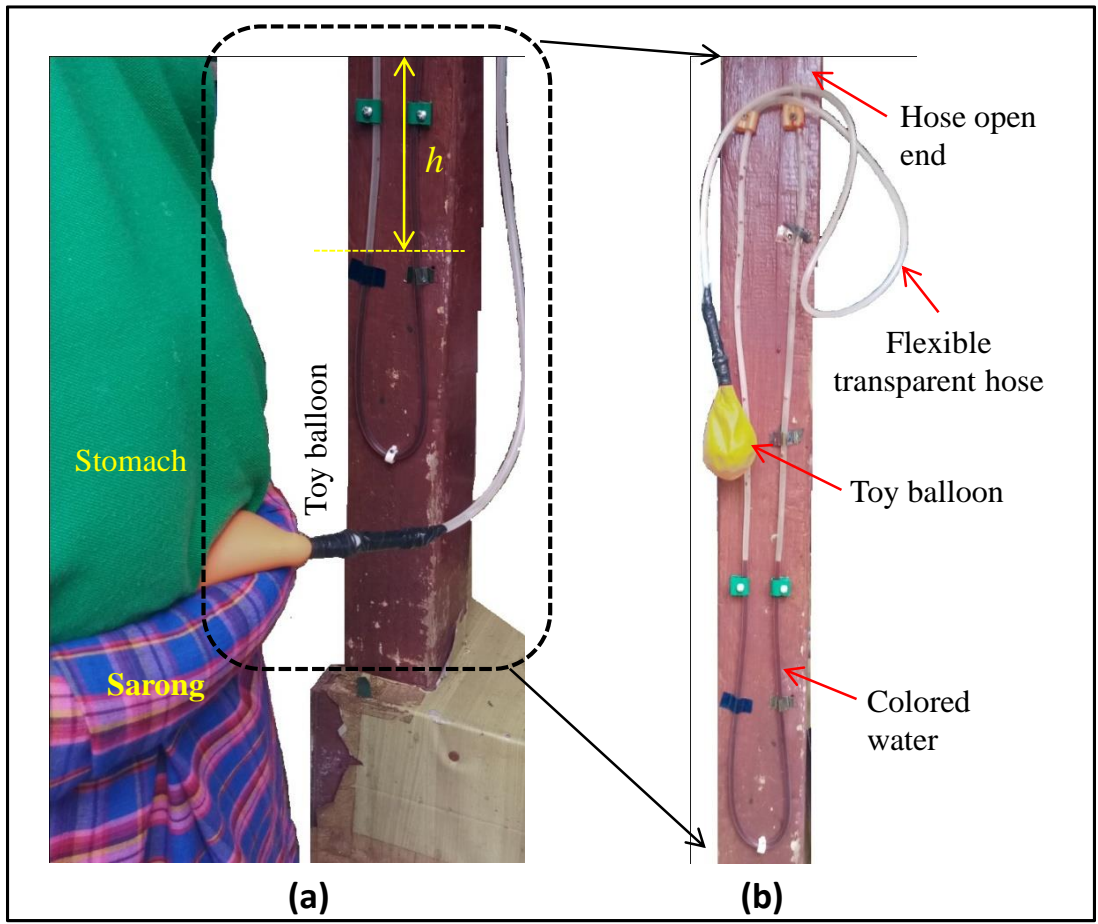

**Figure 1** (a) Measuring the pressure between the stomach and the *sarong*. A toy balloon is inserted between the stomach and the rolled part of the *sarong*, resulting in an increase in the air pressure inside the balloon. It then changed the level of water inside the flexible hose. (b) The detailed image of the system used to measure the air pressure inside the balloon.

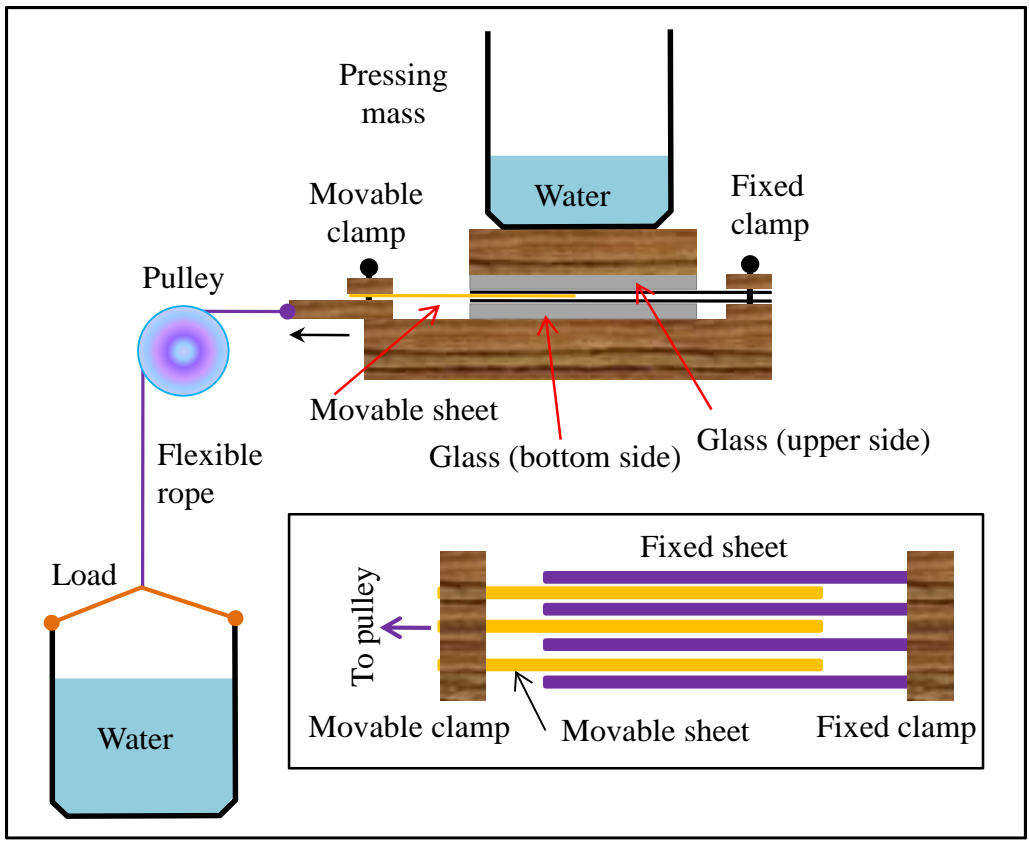

**Figure 2** Equipment for measuring the friction force between the sheets in the stack structure. Inset is the arrangement of the sheets in measurement.

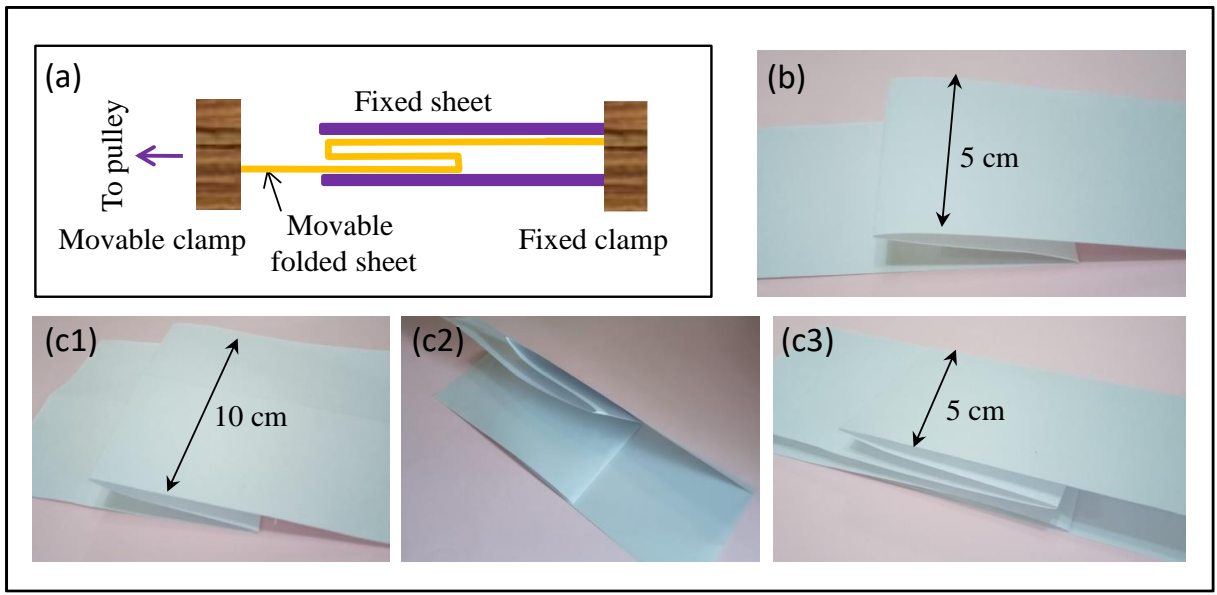

**Figure 3** (a) The arrangement of measuring the force required to release the folding sheets. (b) sheet with zero side folding. (c1) to (c3) processes of making sheet having one side-folding.

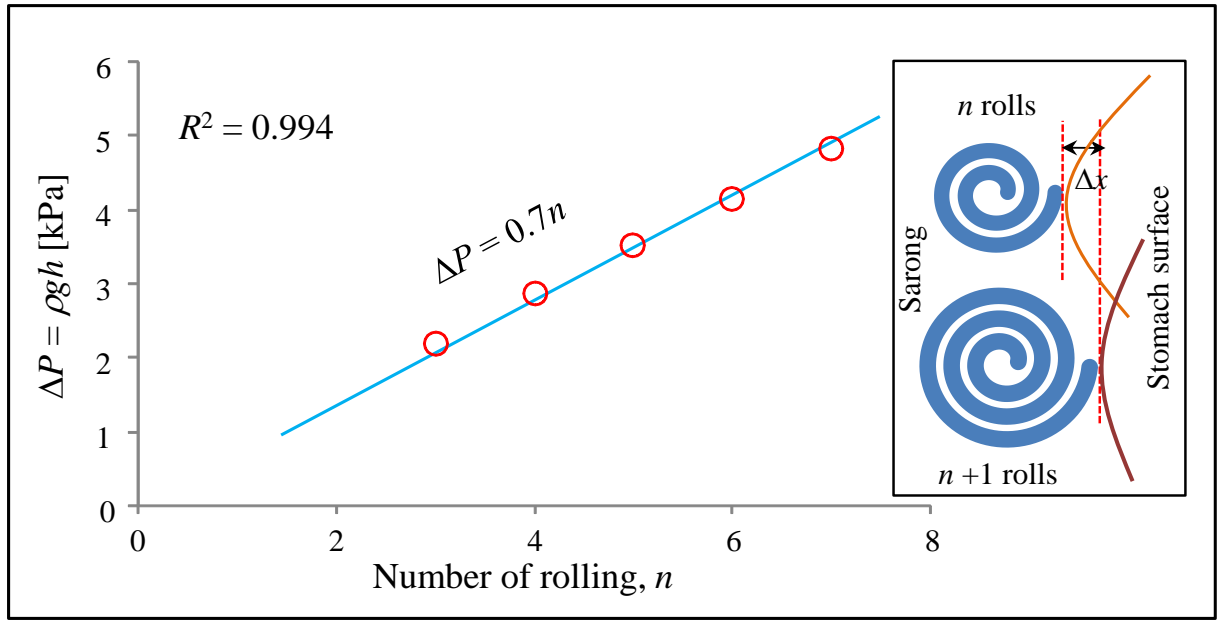

**Figure 4** Effect of the number of rolling the *sarong* on the pressure between the *sarong* and the stomach. Symbols are the measured results and the line is the fitting curve. Inset is the illustration of compression of the stomach surface as the rolling number is increased. $\Delta x$ represents the effective thickness of the *sarong* layer.

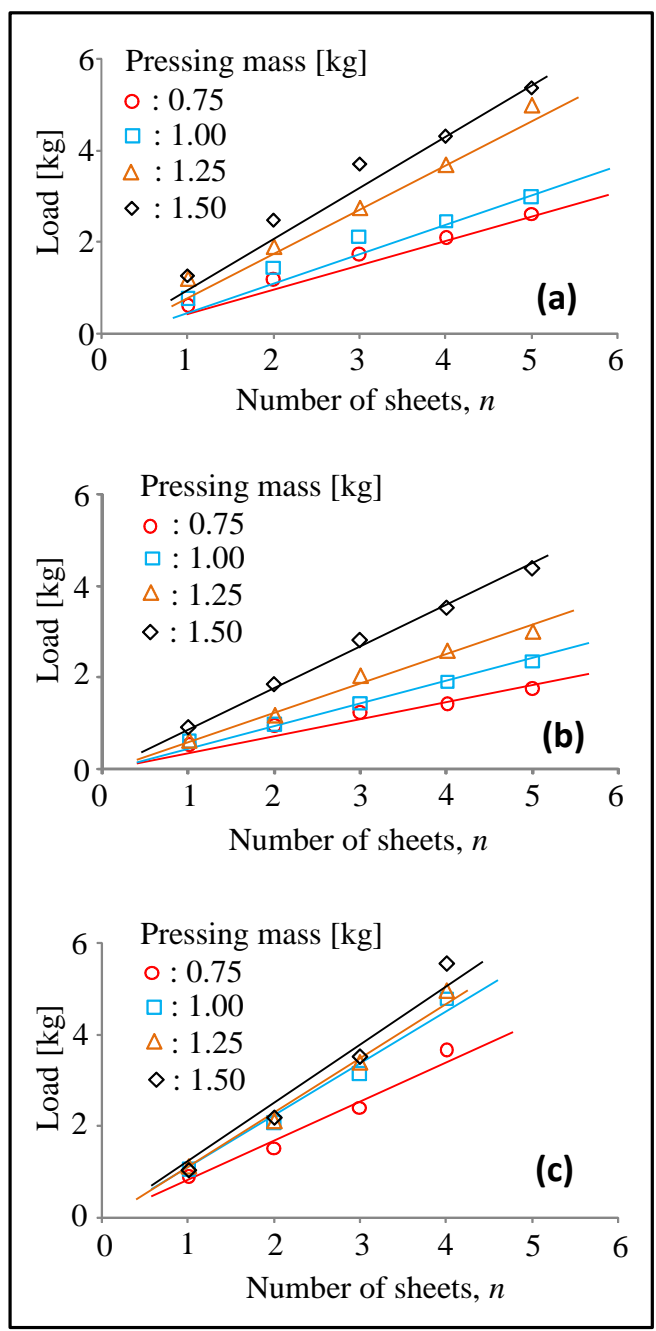

**Figure 5** Effect of the number of movable sheets on the load. We tried three kinds of sheets: (a) photocopy paper sheet, (b) envelope paper sheet, and (c) wool fabric. Four pressing masses were applied to each sheet: 0.75 kg, 1.00 kg, 1.25 kg, and 1.50 kg.

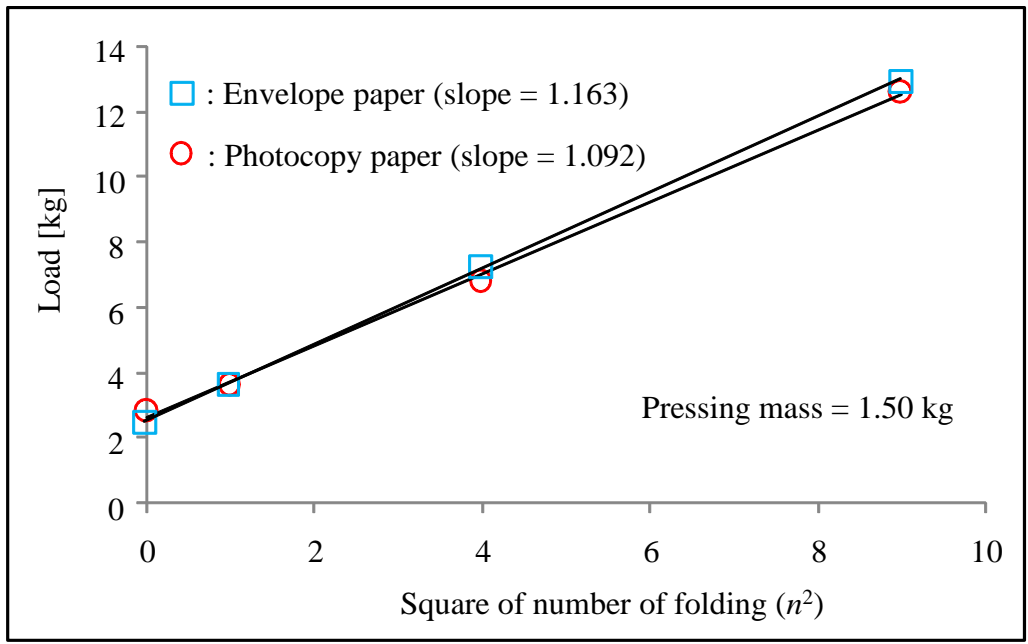

**Figure 6** Effect of squared number of side-folding on the load to start the movement of the sheets: (square) is for the envelope paper sheets and (circle) for the photocopy paper sheets. Symbols are the measured data and lines ate the liner fitting curves. The pressure load was fixed at 1.50 kg.

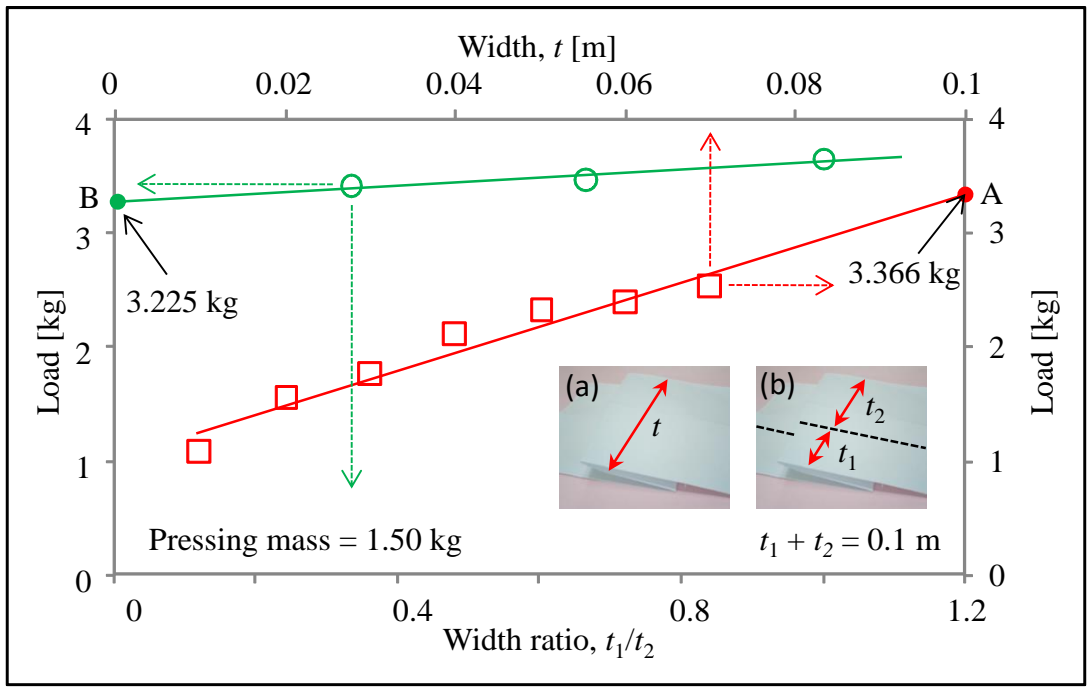

**Figure 7** (circles) The measured data on the effect of folded paper width on the load (the paper was not side folded). The geometry is shown in the inset (a). (square) The effect of asymmetrical side folding width on the load. The geometry is shown in the inset (b). In both experiments, the pressure load was 1.50 kg.

# Appendix

How to wear *sarong*

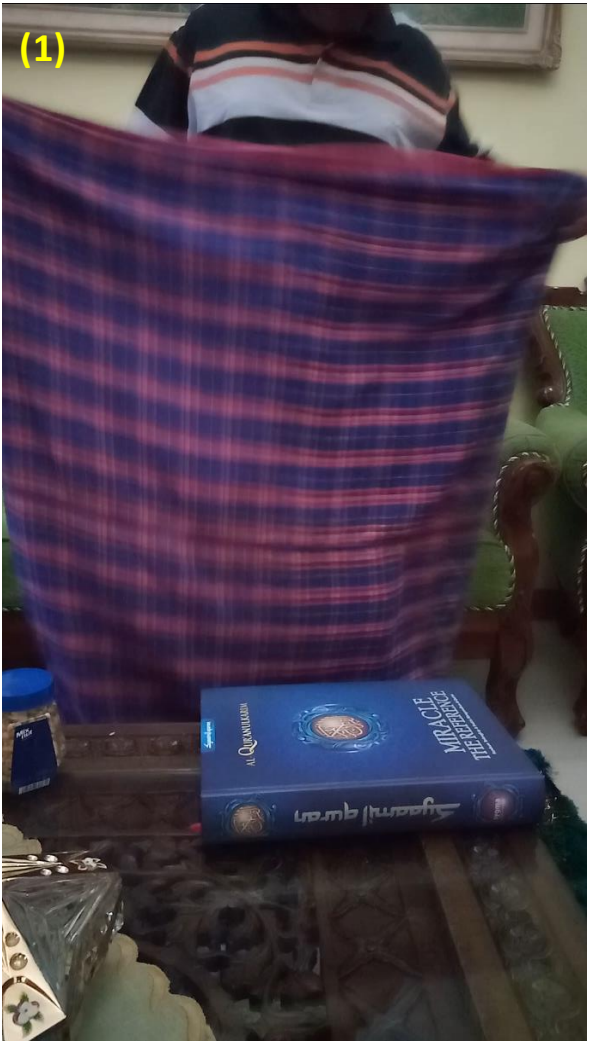 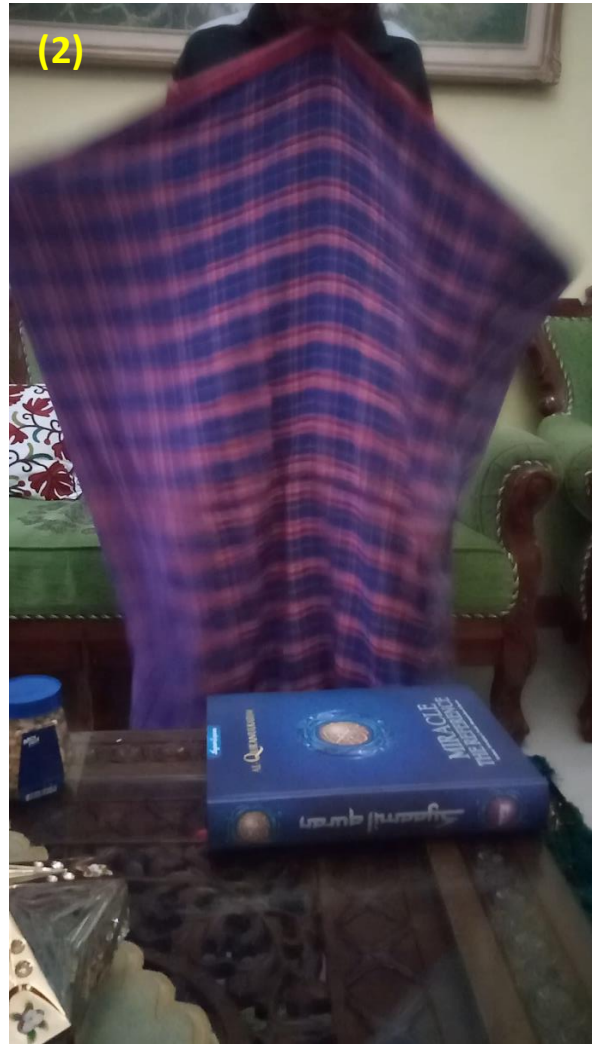

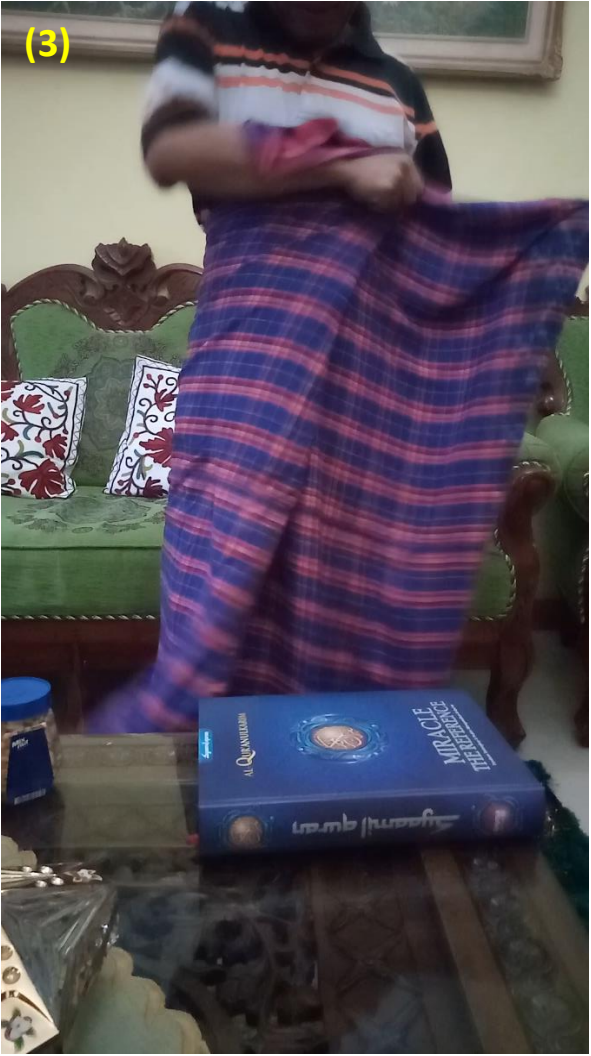 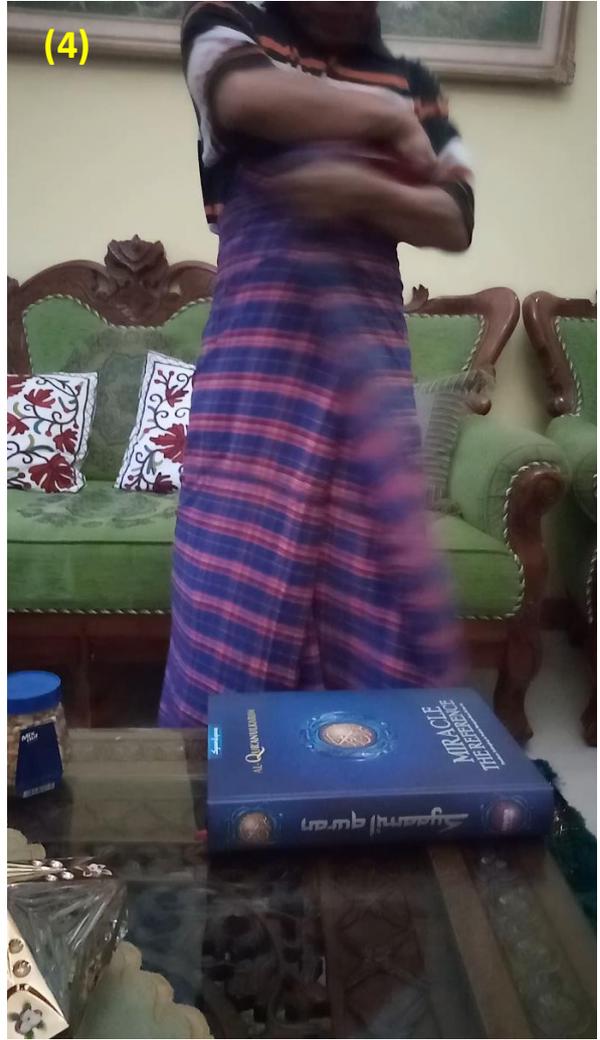

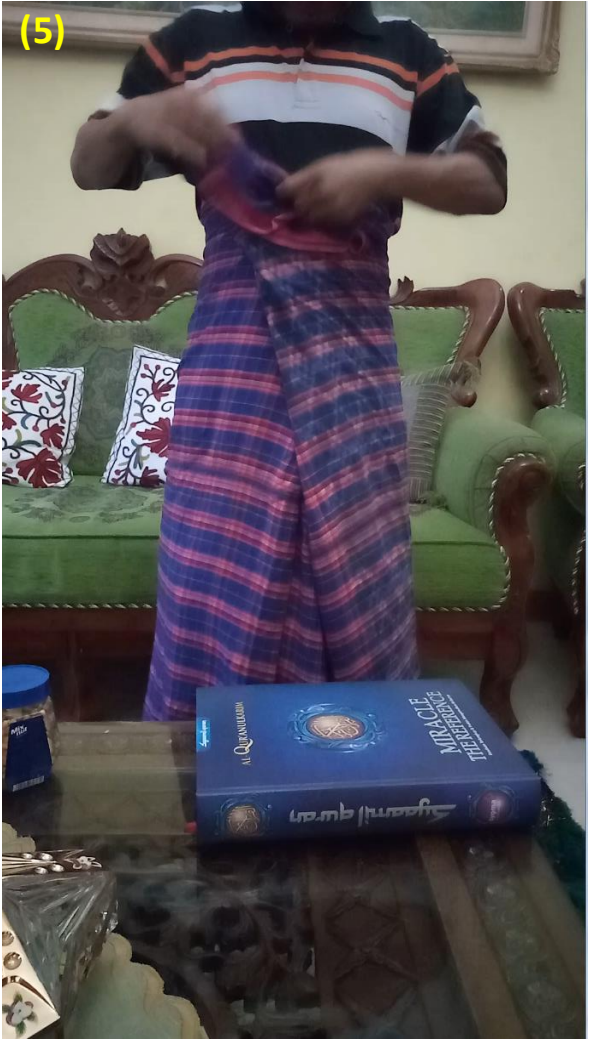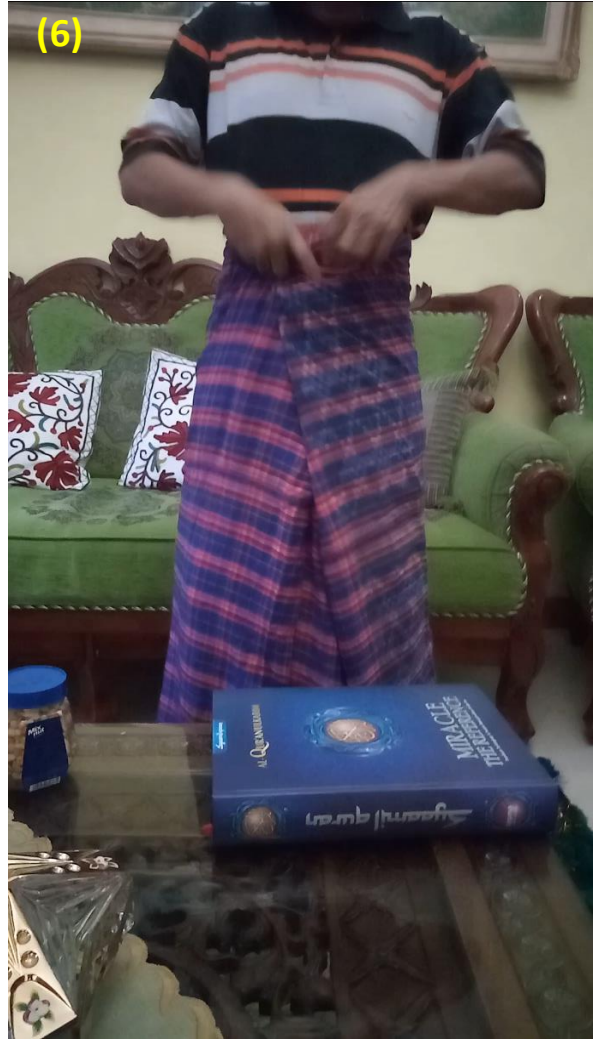

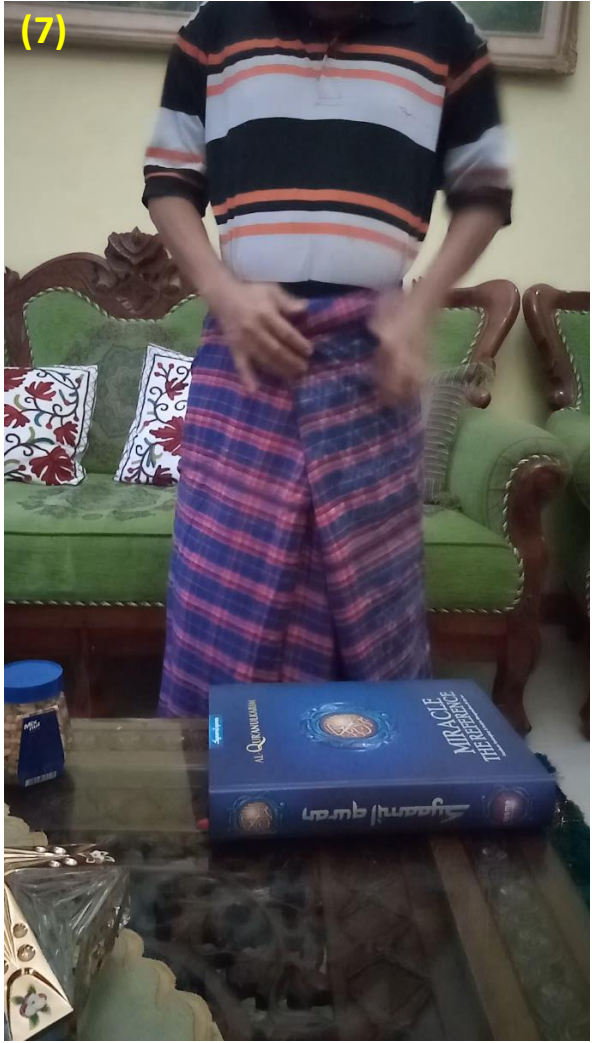 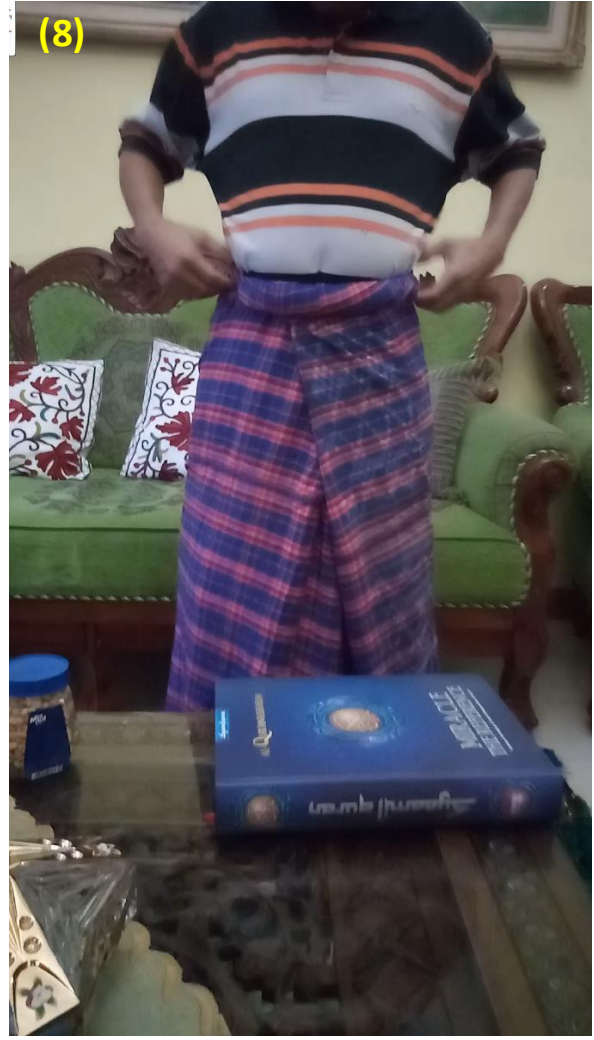

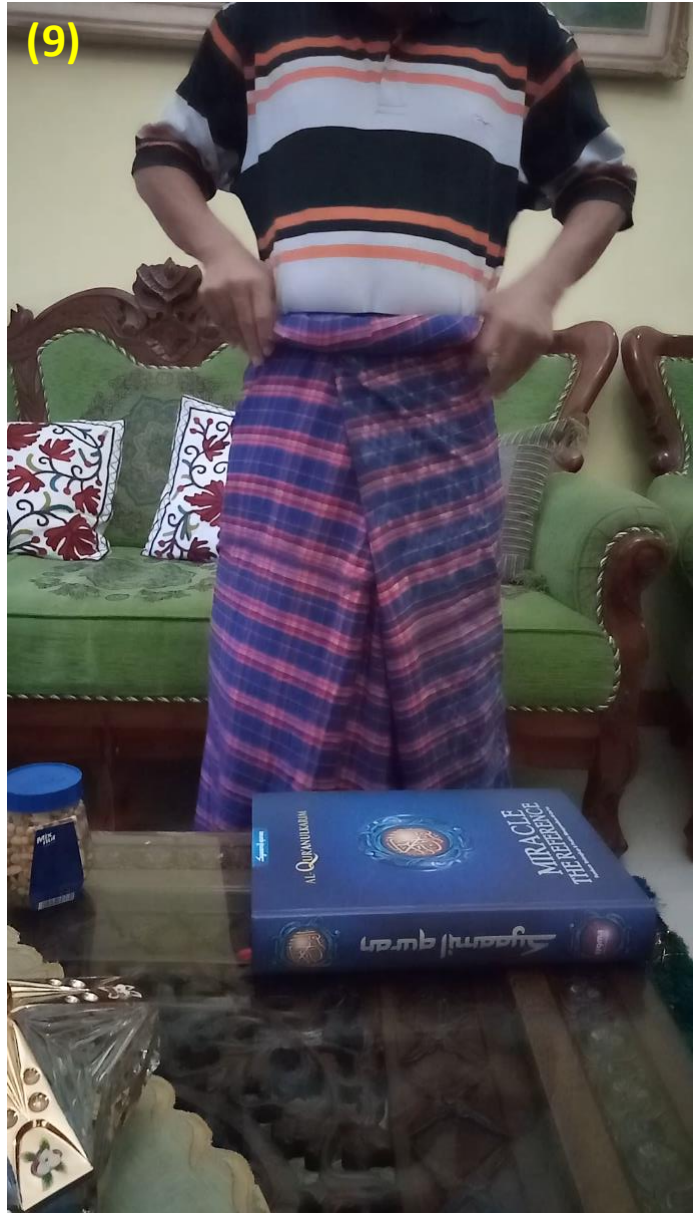